\documentclass[lettersize,journal]{IEEEtran}
\usepackage[letterpaper,margin=1.45cm,footskip=0.4cm]{geometry}
\IEEEoverridecommandlockouts
\usepackage{graphicx}
\usepackage{times,amsmath,amsfonts}
\usepackage{amsmath,amssymb,amstext}
\usepackage{makecell}
\usepackage{ragged2e}
\usepackage{multirow}
\usepackage{footmisc}
\usepackage{tabularray}
\usepackage{hhline}
\usepackage{cite}
\usepackage{epstopdf}
\usepackage[utf8]{inputenc}
\usepackage{bm}
\usepackage{algorithm} 
\usepackage{algorithmicx}
\usepackage{algpseudocode}
\usepackage{color}
\usepackage[lofdepth,lotdepth]{subfig}
\usepackage{soul}
\usepackage{gensymb}
\usepackage{float}
\usepackage{balance}
\usepackage{caption}
\usepackage{subcaption}
\usepackage[inkscapelatex=false]{svg}

\newcommand{\yvec}{\mbox{\boldmath $y$}}
\newcommand{\hvec}{\mbox{\boldmath $h$}}

\newcommand{\xvec}{\mbox{\boldmath $x$}}
\newcommand{\wvec}{\mbox{\boldmath $w$}}

\begin{document}

\title{Digital Twin Enabled Site Specific Channel Precoding: Over the Air CIR Inference \thanks{M. Haider, I. Ahmed, and D. B. Rawat are with the Department of EECS, Howard University, Washington, DC, USA (imtiaz.ahmed@, danda.rawat @)howard.edu. M. Z. Hassan is with the Department of Electrical and Computer Engineering, Universit\'e Laval, QC, CA (md-zoheb.hassan@gel.ulaval.ca). T. O'Shea is with DeepSig Inc., VA, USA (tim@deepsig.ai). L. Liu is with Wireless@VT, Bradley Department of ECE, Virginia Tech, VA, USA (ljliu@vt.edu). (Corresponding author's e-mail: majumder.haider@bison.howard.edu).}}


\author{Majumder Haider, ~\IEEEmembership{Graduate Student Member,~IEEE}, Imtiaz Ahmed, ~\IEEEmembership{ Senior Member,~IEEE}, Zoheb Hassan, ~\IEEEmembership{ Member,~IEEE}, Timothy J. O'Shea, ~\IEEEmembership{ Fellow,~IEEE}, Lingjia Liu, ~\IEEEmembership{ Fellow,~IEEE}, and Danda B. Rawat, ~\IEEEmembership{ Senior Member,~IEEE}
}

\maketitle

\begin{abstract}
This paper investigates the significance of designing a reliable, intelligent, and true physical environment-aware precoding scheme by leveraging an accurately designed channel twin model to obtain realistic channel state information (CSI) for cellular communication systems. Specifically, we propose a fine-tuned multi-step channel twin design process that can render CSI very close to the CSI of the actual environment. After generating a precise CSI, we execute precoding using the obtained CSI at the transmitter end. We demonstrate a two-step parameters' tuning approach to design channel twin by ray tracing (RT) emulation, then further fine-tuning of CSI by employing an artificial intelligence (AI) based algorithm can significantly reduce the gap between actual CSI and the fine-tuned digital twin (DT) rendered CSI. The simulation results show the effectiveness of the proposed novel approach in designing a true physical environment-aware channel twin model.

\end{abstract}


\section{Introduction}
The sixth-generation (6G) of wireless cellular technologies is envisioned to deliver ultra-high spectral efficiency and extremely low latency as well as support huge connectivity for the Internet of Everything (IoE) applications facilitating high-fidelity immersive wireless experiences \cite{han2020artificial},  \cite{serodio20236g}. 
6G seeks to bring revolutionary technologies to ensure full integration of the digital, physical, and human realms for seamless interaction among them. Ultra-massive multiple-input multiple-output (MIMO) will be the key transmission technology for dense environments in 6G. It will employ highly directional narrow beams that can dynamically adapt to changing environments. Efficient precoding at the transmitter end is essential to implement advanced beamforming that can track users in real-time, notably enhancing overall system efficiency. Since precoding is a function of channel state information (CSI), to facilitate precoding, precise knowledge of CSI with low overhead or without overhead at the transmitter end is highly significant. 

However, having perfect knowledge of CSI at the transmitter end with low overhead is very challenging. Conventionally, channel estimation in predefined pilot carrying time slots at the receiver end and feeding it back to the transmitter to obtain CSI increases computational overhead. 
On the other hand, now-a-days deep learning (DL) aided CSI acquisition technique is receiving significant attention as an alternative approach providing the benefits of offline learning and online implementation \cite{soltani2019deep, yi2020deep} to reduce overhead significantly. Nevertheless, CSI depends on the complex and dynamic physical environment, which is also site-specific. It is difficult to appropriately learn or model such dependencies with low complexity using the DL technique solely, even with a large volume of site-specific data, a data-driven strategy at the transmitter may not always effectively capture site-specific CSI.

A virtual version of a physical end-to-end communication system that replicates its behavior, environment, and performance in real-time is known as a digital twin (DT) \cite{alkhateeb2023real, ruah2024calibrating, zhu2024digital}. 
DT generates context-aware datasets for a well-defined network to be used by the artificial intelligence (AI) / machine learning (ML) model to predict near-realistic network behavior that targets to reduce the computational overhead of the real networks, as well as DT can reduce the necessity of the control channel signaling, hence more time slots can be utilized for data transmission to ensure bandwidth efficiency.

To address the challenge of pilotless CSI acquisition for the entire transmission in practical communication systems, we propose a novel approach to designing a channel twin model that facilitates intelligent precoding integrating the concepts of DT utilizing a locally calibrated ray tracing (RT) model. The paper \cite{cao2024channel} discussed the visions of DT-aided channel twining for CSI acquisition, but no experiment and no specific approach are provided to achieve the channel twin model. In \cite{10622316}, although the DT-aided CSI compression and feedback technique was illustrated to obtain CSI close to the real CSI, it did not highlight the impact of the tuning approach of the RT parameters, nor did it use real CSI in the training phase to effectively learn the DL model. In contrast, in this paper, we demonstrate the detailed experimental analysis used to design the DT-assisted true environment-aware channel twin model to achieve CSI that closely matches the real CSI. In particular, our work emphasizes designing a truly physical environment-aware precoding integrated site-specific channel twin model of the corresponding physical scenario. The developed precoding approach uses the DT-aided fine-tuned channel impulse response (CIR) to notably reduce the computational overhead of computing CSI conventionally. The steps and specifications proposed to design the channel twin model can provide an intelligent and efficient approach that can efficiently replicate the physical real-world CSI. 
The comparative performance evaluation utilizing CIR in terms of bit error rate (BER) demonstrates the effectiveness and accuracy of the proposed novel channel twin design approach.

\section{System Model}
\noindent\textbf{Signal Model:} Fig. 1 shows a multi-input single-output orthogonal frequency division multiplexing (MISO-OFDM) downlink communication system. We consider the time-varying frequency-selective multipath fading channel along with the deployment of linear precoding to maximize the received signal power by exploiting transmit diversity. The base station (BS) is equipped with $N_{T}$ antennas to create transmit diversity and a single antenna at the user equipment (UE) for the signal reception. The system operates over a transmission bandwidth divided into $N_{s}$ orthogonal subcarriers. Let us assume that the transmitted information bits are represented by $s$, the corresponding baseband modulated symbols (e.g., binary phase shift keying (BPSK), quadrature phase shift keying (QPSK), $M$-ary quadrature amplitude modulation ($M$-QAM), etc.) for the $k\in\{1,2,..., N_{s}\}$  OFDM subcarrier are denoted as $S_{k}$. 
$B_{k}$ represents the precoding factor, which is a function of the underlying channel between BS and UE, depending on the type of precoding scheme. The precoded transmitted signal for antenna $t \in \{1,2,\cdots,N_T\}$ can be expressed as $X_{k}^t = S_{k}B_{k}^t$. Let us denote $H_{k}^t$ as the underlying frequency domain channel response for subcarrier $k$ and the communication link between antenna $t \in \{1,2,\cdots,N_T\}$ of BS and UE. After performing the inverse fast Fourier transform (IFFT), the time domain representation of $X_k^t$ and the channel impulse response (CIR) can be denoted as $x_{n_1}^t[m]$ and $h_{n_2}^t[m]$, respectively, for the OFDM symbol $m = \{1,2,\cdots\}$. Note that $n_1 \in \{1,2,\cdots,N_s\}$ and $n_2 \in \{1,2,\cdots,\eta_t\}$ present the indices for time domain signal and CIR, respectively. Note that $\eta_t$ denotes the length of (frequency-selective) CIR for the communication link between antenna $t \in \{1,2,\cdots,N_T\}$ of the BS and the UE. Adding the cyclic prefix (CP) of $\tau$ samples to each OFDM symbol $m$ yields $\Tilde{x}_n^t[m]$, $n \in \{1,2,\cdots,N_s + \tau\}$. 
\begin{figure}[t!]
\centering
\includegraphics[width = 8cm, height = 5cm]{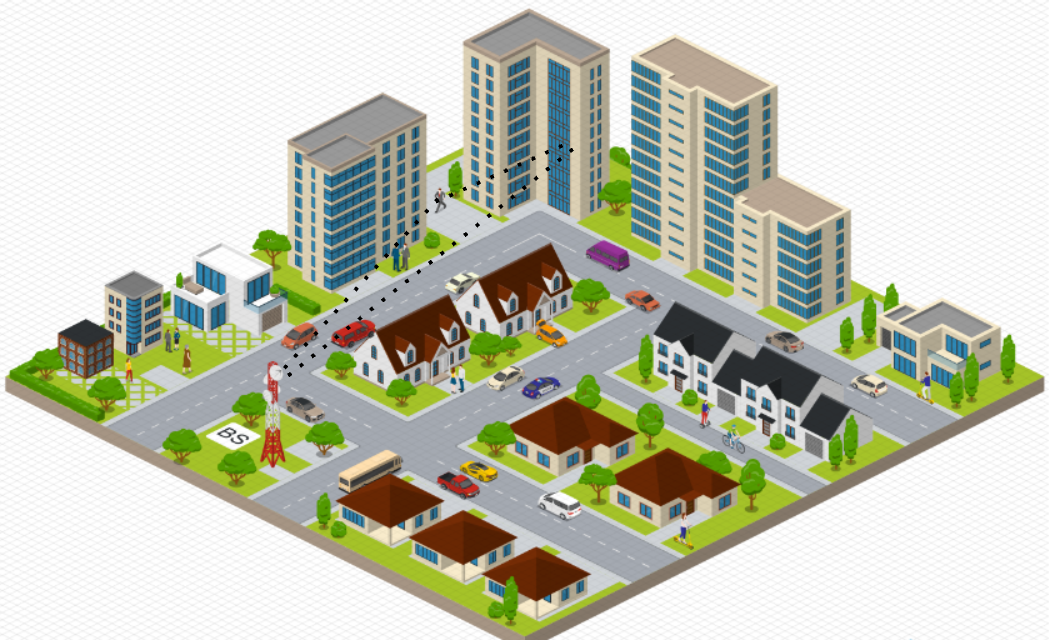}
\captionsetup{justification=centering}
\caption{Downlink MISO communication system.}
\label{fig1}
\end{figure}
The received signal vector for OFDM symbol $m$ from $N_T$ transmit antennas can be expressed as 
\begin{equation}
 \yvec[m]= \sum_{t=1}^{N_T} \Tilde{\xvec}^t[m] \circledast \hvec^t[m] + \wvec[m], 
 \label{eq1}
 \end{equation}
where  $(\circledast)$ represents convolution operation, $\wvec[m]$ denotes the additive white Gaussian Noise (AWGN) vector, where each element possesses zero mean and variance $\sigma_w^2$. Here, $\Tilde{\xvec}^t[m]$ and $\hvec^t[m]$ present the transmit signal vector of size $1 \times (N_s + \tau)$ and CIR of size $1 \times \eta_t$, respectively for OFDM symbol $m \in \{1,2,\cdots\}$. 
UE removes the CP and executes the fast Fourier transform (FFT) operation to obtain the signals in the frequency domain. Without loss of generalization, we consider perfect channel knowledge for data detection. In practice, different blind estimation techniques can be applied at the UE for estimating the CIR and can thereby eliminate the requirements of pilot transmission, albeit at the expense of higher computational complexity. The transmit signal can be detected at the receiver by applying linear and non-linear equalization schemes. The equalized signal is expressed as $\hat{S}_k$, which is then demodulated to obtain the estimated bits $\hat{s}$.

\noindent\textbf{Precoding Schemes:} We consider two linear precoding schemes: maximum ratio transmission (MRT) and minimum mean square error (MMSE). For MRT precoding, the precoding vector for antenna $t$,  $B_k^t$ is computed as $B_k^t = (\hat{H}_k^t)^{*}$ \cite{selvan2014performance}, where $(\ast)$ represents the conjugate operation and $\hat{H}_k^t$ is the estimated channel frequency response. It is worth mentioning that the objective of this paper is to obtain $\hat{H}_k^t$ from the proposed true environment-aware channel twin model. For MMSE precoding, $B_k^t = (\hat{H}_k^t)^{*} ((\hat{H}_k^t)^{T} (\hat{H}_k^t)^{*} + 1 / {P_t})$ \cite{selvan2014performance}, where $P_t$ is the transmit power for antenna $t \in \{1,2,\cdots,N_T\}$. Here, $(\cdot)^T$ denotes the transpose operation. 

\section{Environment Aware Channel Twin Model}
In this section, we describe our proposed multi-step channel twin design process including the real-world CIR data collection. The architecture and the functionality of the proposed AI-driven channel twin model are presented. 

\subsection{Real Radio Environment}
\begin{figure}[h!]
\centering
\includegraphics[width = 6cm, height = 6cm]{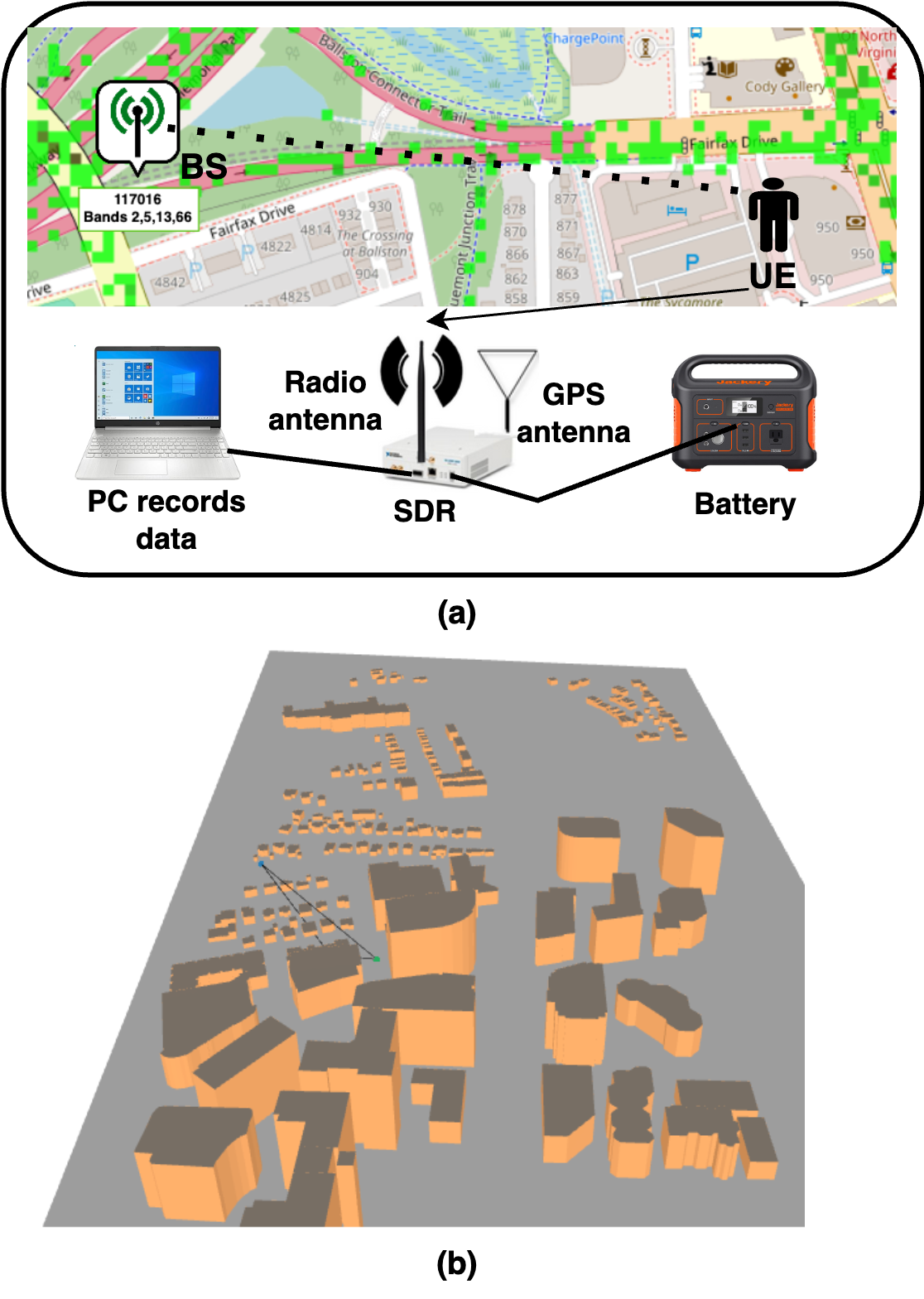}
\captionsetup{justification=centering}
\caption{(a) Real data measurement (b) 3D DT model.}
\label{fig2}
\end{figure}
In Fig. \ref{fig2}(a), we depict the location of the actual real-world environment where we collected CIR. We captured the demodulation reference signal (DMRS) from the BS with GPS coordinates latitude 38.88175\degree and longitude -77.12180\degree, through a driving test along North Glebe Road in Ballston, Arlington, Virginia, USA. Moreover, Fig. \ref{fig2}(a) includes the pictorial view of the software-defined radio (SDR) setup used to capture real-time field CIR measurement data using DeepSig's hcapture software. We used NI USRP B210 as the radio receiver integrated with a GPS antenna and a single radio antenna. The mean GPS coordinates of the receiver are latitude 38.88131\degree and longitude -77.11655\degree. The numerical values of the cell tower parameters' are illustrated in Table \ref{tab1}.
\setlength{\textfloatsep}{0pt}
\begin{table}[h]
\centering
\captionsetup{justification=centering}
\caption{Cell tower parameters}
\begin{tabular}{|p{2.3cm}|p{3.5cm}|}
\hline
\bfseries{Parameter} & \bfseries{Value}  \\
\hline
Channel Bandwidth & 10MHz \\
\hline
Uplink Frequency & 782MHz  \\
\hline
Downlink Frequency & 751MHz  \\
\hline
Frequency Band & B13 FDD (n13 in 5G NR)  \\
\hline
\end{tabular}
\label{tab1}
\end{table}

\subsection{Ray Tracing Aided 3D Modeling}
In this subsection, we emphasize RT-aided three-dimensional (3D) modeling to create an initial version of a channel twin model, which is then tuned with an optimized material selection
process to create a digital replica of the real-world wireless environment. In particular, we demonstrate this two-step tuning approach of the RT 3D model to render high-fidelity (HF) CIR. 

\subsubsection{\textbf{RT Propagation Physics Optimization}}
We portrayed a virtual 3D environment of the location of our interest shown in Fig. 3(b). First, we import the specific location (the place where real data was captured) of Ballston, Arlington, Virginia using OpenStreetMap into Blender. We then add the ground plane and scale the imported 3D map in Blender. In the 3D model, all buildings and roads are defined with the appropriate radio materials. The model is then imported into the SIONNA-RT \cite{hoydis2023sionna} so that the RT tool can identify the objects of the 3D model for wireless signal multipath propagation. After importing the 3D virtual model of the real environment into SIONNA, we incorporate a transmitter (BS) and a receiver (UE) into the model by defining their location, antenna parameters, and operating frequency. A key feature of RT is the handling simulation of multipath propagation, where a signal from the transmitter reaches the receiver through multiple paths such as line of sight (LoS) path, reflected, diffracted, and scattered paths due to the propagation physics. In the RT emulation, we consider the impact of all three (reflection, diffraction, scattering) major propagation physics to incorporate the characteristics of the real environment. Moreover, the number of rays traced has a crucial impact on generating realistic CIR. Since an increasing number of rays can capture multipath signal propagation effects and the signal interaction with the surface and objects more precisely, it can thus render higher accuracy in creating a channel twin model and provide high-fidelity predictions of signal strength and coverage, optimizing the parameters pertinent to propagation physics. We have illustrated the comparative analysis of varying numbers of ray impact in generating CIR in Section IV. However, the interaction of a large number of rays in complex environments is computationally resource intensive. It requires significant computational power and time to execute the simulation while incorporating a large number of rays into the 3D RT model.

\subsubsection{\textbf{EM Material Optimization}}
In RT emulation, the use of electromagnetic (EM) materials plays a crucial role in how electromagnetic waves interact with different surfaces and objects such as buildings, roads, and terrains. The interaction of rays with EM materials depends on their EM properties, such as refractive index, permittivity, conductivity, absorption, and scattering properties. To make 3D RT emulation more realistic, complex calculations are required to properly handle the interaction of rays with EM materials, especially when multiple effects occur simultaneously (e.g., reflection, diffraction, and scattering). Moreover, different types of EM material effects influence how rays reflect, diffract, and scatter when interacting with an EM material \cite{series2015effects}. 
By combining these effects, RT algorithms can produce highly realistic digital replicas that mimic real-world physical radio signal interactions.
Anticipating that fact, we implement the following 3 different frequently used EM material choices (MC) described in Table \ref{tab2} as trial and error 
\setlength{\textfloatsep}{0pt}
\begin{table}[h]
\centering
\captionsetup{justification=centering}
\caption{Different EM materials}
\begin{tabular}{|p{0.99cm}|p{2.2cm}|p{0.99cm}|p{0.99cm}|}
\hline
\bfseries{Category} & \bfseries{Ground Plane \& Roads} & \bfseries{Building Wall} & \bfseries{Building Roof}  \\
\hline
MC1 & itu\_concrete & itu\_marble & itu\_metal  \\
\hline
MC2 & itu\_concrete & itu\_marble & itu\_wood  \\
\hline
MC3 & itu\_concrete & itu\_brick & itu\_wood  \\
\hline
\end{tabular}
\label{tab2}
\end{table}
basis to determine the most suitable EM material that can render higher accuracy in RT emulation to reduce the gap between actual CIR and the RT rendering CIR. We have shown the comparative effect of different EM materials in the generation of CIR in Section IV.

To further optimize the RT parameters and thus generate a close-to-actual CIR, we propose an AI-driven scheme that learns the correlation between actual and RT-generated HF CIRs \cite{ahmed2024integrated}. In the next subsection, we describe the architecture and functionality of the proposed AI-driven algorithm.

\subsection{Architecture AND Functionality of AI-Driven Algorithm}
\begin{figure}[h]
\centering
\includegraphics[width = 7cm, height = 6cm]{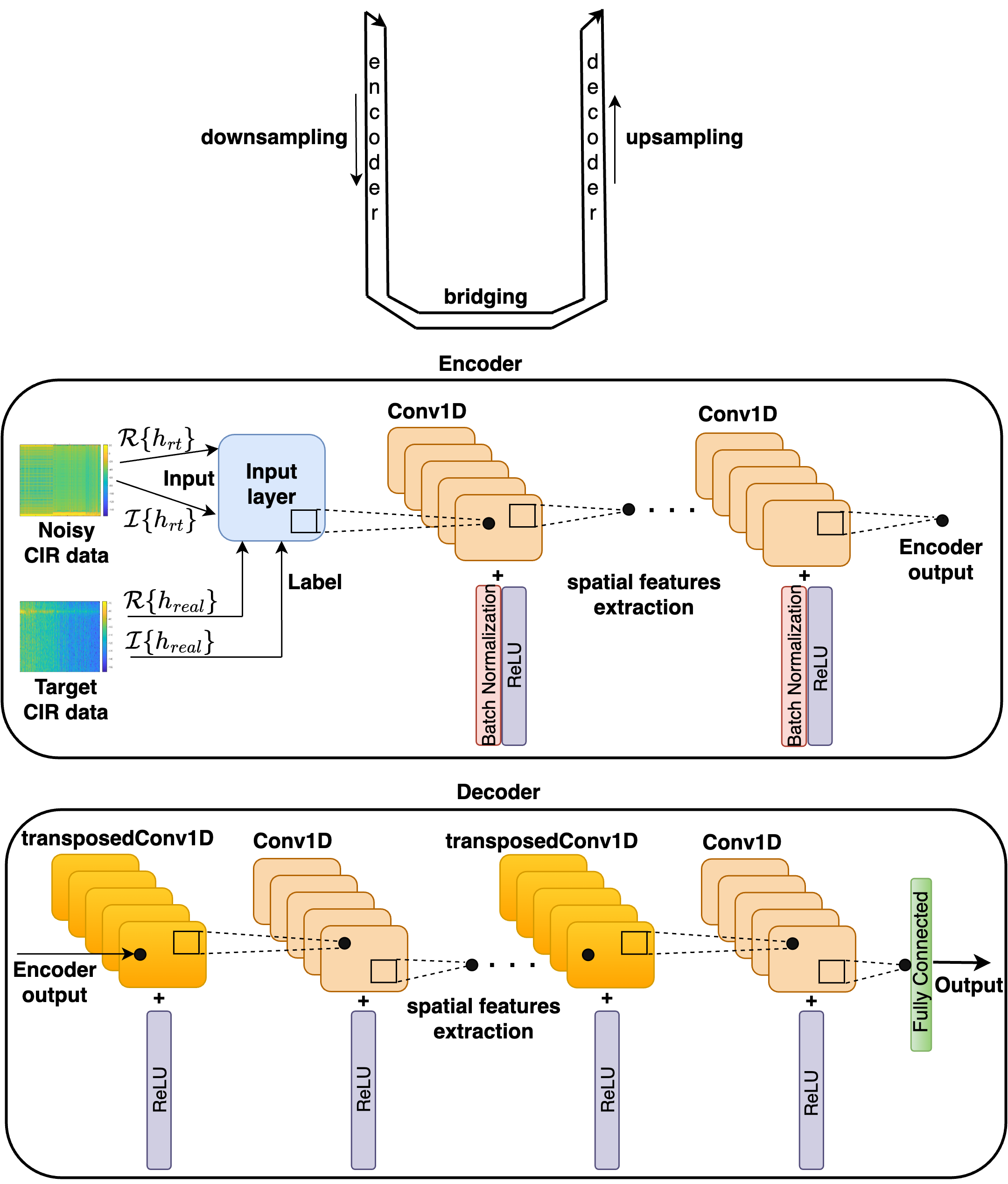}
\captionsetup{justification=centering}
\caption{Neural network model architecture.}
\label{fig3}
\end{figure}

The proposed AI-driven algorithm incorporates the modified U-Net architecture, which consists of two main parts: a compression path called the encoder and an expansion path called the decoder.
The deep denoising U-Net model with encoder and decoder architecture was proposed with very high precision for the image segmentation task \cite{ronneberger2015u}.
The primary objective of our AI-driven algorithm is to analyze the correlation between the synthetic CIR dataset from the RT emulation after applying the aforementioned two-step tuning approach and the actual CIR captured through the driving test. To facilitate the objective, refer to Fig. \ref{fig3}, the encoder neural network (NN) consists of an input layer and $\mathcal{L}_{e}-1$ hidden blocks. The input features of the encoder NN are $\mathcal{R}$$\{h_{rt}\}$ and $\mathcal{I}$$\{h_{rt}\}$, where $\mathcal{R}$ and $\mathcal{I}$ denote the real 
and imaginary parts, respectively, of the noisy HF CIR generated in RT emulation after executing a two-step tuning approach, and the corresponding output labels of the encoder NN model are $\mathcal{R}$$\{h_{real}\}$ and $\mathcal{I}$$\{h_{real}\}$. The dimension of the raw CIR dataset is ($\mathcal{S}$ $\times$ $\mathcal{A}$ $\times$ $\mathcal{P}$), where $\mathcal{S}$, $\mathcal{A}$, and $\mathcal{P}$ represent the numbers of samples, transmit antennas, and paths, respectively. After reshaping the CIR dataset, the dimension of the input features can be represented as ($\mathbb{R}$ $\times$ $\mathbb{F}$), where $\mathbb{R}$ and $\mathbb{F}$ represent the number of realizations and input features (real and imaginary of a complex variable), respectively. Each $\mathcal{L}_{e}-1$ hidden block is composed of the single-dimensional convolution (Conv1D) layer along with varying numbers of kernels and kernel size followed by the batch normalization layer to ensure stable learning and the rectified linear unit (ReLU) activation layer to integrate non-linearity. The objective of hidden blocks is to extract spatial features that execute the compression of the input data.
On the other hand, the decoder NN consists of $\mathcal{L}_{d}-1$ hidden blocks. Each $\mathcal{L}_{d}-1$ hidden block is comprised of the transposed Conv1D layer with the ReLU activation layer followed by the Conv1D layer with the ReLU activation layer. The number of kernels and the kernel size vary for each block (tuned via trial-and-error basis). The goal of the hidden blocks is to extract spatial features by upsampling the compressed data processed and forwarded by the encoder NN to reconstruct the transmitted signal and its original dimension. The final layer is the fully connected output regression layer. 
We denote $u\in\{1,2,..., U\}$ and $v \in\{1,2,..., V\}$, where $U$ and $V$ as the total numbers of epoch for training and the total numbers of batches in each training epoch, respectively.
The parameters of the deep NN (DNN) are defined as $\bm{\alpha}_g$ and $\bm{\beta}_g$ for a given hidden-layer $g \in \{1,2,\cdots,{G}\}$, where  $\bm{\alpha}_g$ and $\bm{\beta}_g$ denote the weights and bias factors, respectively. The estimation of the predicted CIR $\hat{\hvec}_{pred}^u$ can be written as 
\begin{equation}
\label{ch_estm}
  \hat{\hvec}_{pred}^u = f_{G}\left(\bm{\alpha}_{G}f_{G-1} \left(\cdots f_{1} \left( \bm{\alpha}_{1} {\hvec_{rt}}^u + \bm{\beta}_{1} \right) \cdots\right) + \bm{\beta}_{G} \right)  
\end{equation}
where $f_g(\cdot)$ is the activation function at the $g$-th hidden layer, $\forall g$. The objective function of the NN model can be mathematically expressed as
\begin{align}
o_{NN}^* = \arg \min _{o_{NN}} \mathbb{E}_{\hat{\hvec}_{pred}^u \mapsto \hvec_{real}^u} \left\{ \left| \left|\hvec_{real}^u-\hat{\hvec}_{pred}^u  \right| \right|^2 \right\}, \label{eq5}
\end{align} 
 where $\mathbb{E} \left\{ \cdot \right\}$ and $\left|\left| \cdot \right|\right|^2$ represent the statistical expectation and $L_2$-norm operations, respectively. The loss function is the mean square error (MSE), and the optimizer is Adam, using a stochastic gradient descent approach with an initial learning rate $\xi$. 

\section{Simulation results}
\noindent\textbf{Parameter Specifications:} In this section, we present the numerical performance evaluation for the proposed multi-step optimization process to provide optimized CIR close to the real CIR sequentially. The performance metric is the bit error rate (BER). 
The effective signal-to-noise ratio (SNR) is computed as $SNR = {1}/{\sigma_{w}^2}$. We assume that a precise computation of channel inversion is conducted. Throughout the simulations, we consider the QPSK modulation scheme and zero-forcing equalizer for data detection.

\begin{figure}[t!]
\centering
\includegraphics[width = 8cm, height = 6cm]{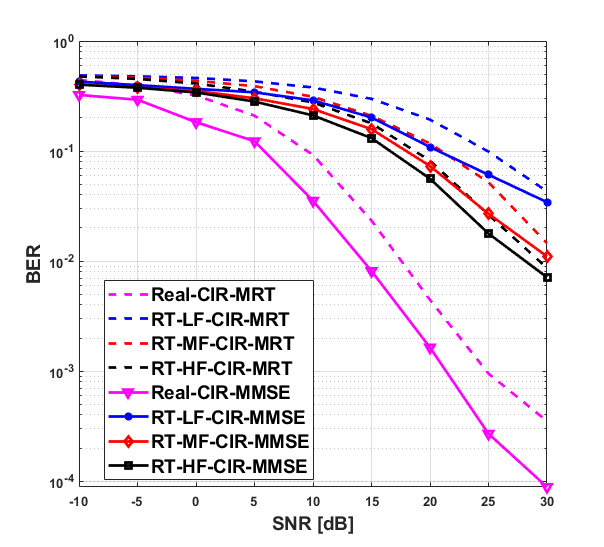}
\captionsetup{justification=centering}
\caption{BER comparison for different precoding prioritizing the impact of the number of rays.}
\label{fig4}
\end{figure}
\begin{figure}[h!]
\centering
\includegraphics[width = 8cm, height = 5.7cm]{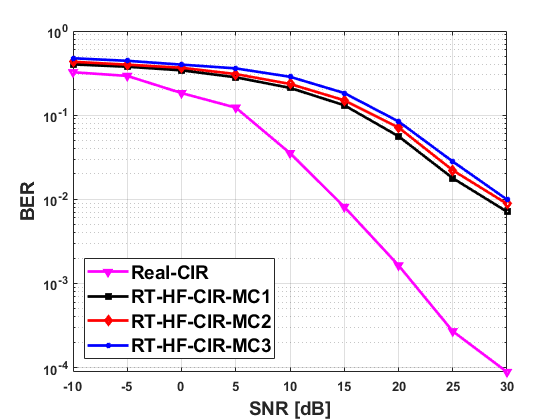}
\captionsetup{justification=centering}
\caption{BER comparison considering the EM material impact.}
\label{fig5}
\end{figure}
\begin{figure}[h!]
\centering
\includegraphics[width = 8cm, height = 5.7cm]{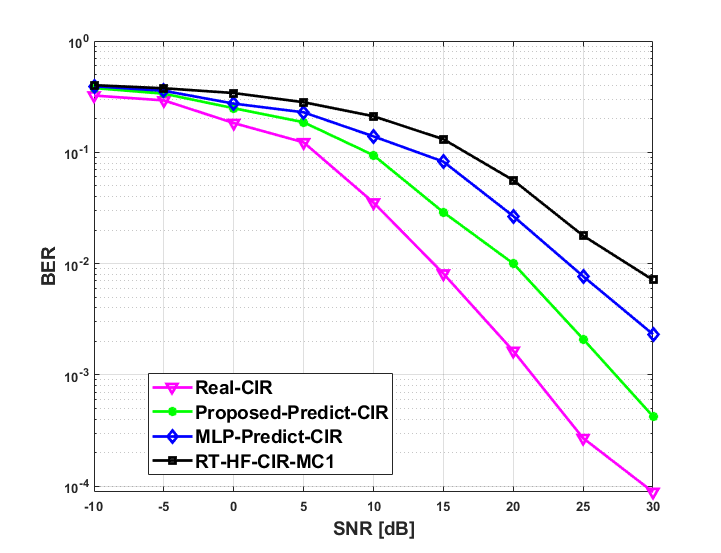}
\captionsetup{justification=centering}
\caption{BER comparison based on real, DT model and NN predicted CIR.}
\label{fig6}
\end{figure}

For the CIR prediction task, noisy and real CIR dataset scaling, dataset separation into training, testing, and the DNN modeling are executed using the MATLAB DL toolbox via Monte Carlo simulations. We assume $N_{T} = 4$, $N_{s} = 128$, $\tau = 16$, $\xi = 0.0008$, $\mathcal{L}_{e} = \mathcal{L}_d = 5$, and $\mathbb{L} = 5$ unless otherwise stated. We consider total $\mathbb{R} = 112,000$ realizations of data samples that are separated into training and testing purposes as $70\%$ and $30\%$, respectively for both the proposed and baseline schemes, and $\mathbb{F} = 2$.
In the DNN-aided prediction task, we consider the same environment for training and testing while conducting simulations for experiments.
It is worth noting here that after importing the RT emulation-generated CIR into MATLAB, we first scale the dataset as well as the real CIR dataset. We consider the sampling frequency $\mathcal{F}_s = 30.72$ MHz. Unless otherwise stated, Fig. \ref{fig4} considers the MRT and the MMSE precoding scheme, Fig. \ref{fig5}, and Fig. \ref{fig6} consider only the MMSE precoding scheme at the BS.

\noindent\textbf{AI-driven Baseline Scheme: }A fully connected multi-layer perceptron (MLP) model is considered as a baseline AI-driven scheme to demonstrate the effectiveness of our proposed scheme. The fully connected DNN model consists of a sequential input layer, $\mathbb{L}-2$ hidden layers, and a final output regression layer.

\noindent\textbf{Genie-Aided Scheme: } In this scheme, we consider that the exact CIR is available at the transmitter for calculating $B_k^t$. Note that this scheme is not practically realizable. However, it results in a lower bound of BER that assists in presenting the relative performances of the proposed scheme.

\noindent\textbf{Performance Evaluation of RT Propagation Physics Optimization:} In Fig. \ref{fig4}, we show the impact of different numbers of rays in generating CIR on the end-to-end BER, computed while utilizing the captured actual CIR (genie-aided scheme) and the CIR generated by the RT emulation. As expected, increasing the number of rays decreases the gap between the BERs computed using actual CIR  and the CIR generated by RT emulation. A large number of rays in RT assists in capturing multipath propagation scenarios effectively. It is observed that the decreasing rate of BER is much higher in the higher range of SNR values when switching from RT-low-fidelity (LF) CIR (number of rays considered $10^3$) to RT-moderate-fidelity (MF) CIR (number of rays considered $10^4$) compared to RT-MF CIR to RT-HF CIR (number of rays considered $10^5$).  Fig. \ref{fig4} also illustrates the performance gap between the MRT and the MMSE linear precoding technique. 

\noindent\textbf{Performance Evaluation of EM Material Optimization:}
Fig. \ref{fig5} illustrates the significance of choosing proper EM material to get reasonable BER performance. Although the BER performance gap among three different EM MC is not significant, the appropriate EM material selection can play an important role in reducing the BER gap calculated using actual CIR (genie-aided precoding) and the HF CIR generated by RT emulation. Refer to Table \ref{tab2}, EM MC1 gives lower BER compared to others in the considered 3D scene.

\noindent\textbf{Performance Evaluation of the Proposed DNN Model Prediction:} In Fig. \ref{fig6}, we show the significant improvement of the BER performance after using the predicted CIR from the comprehensive channel twin, where the RT scenario is enhanced by data-driven approach. The proposed supervised DNN model is trained with the two-step fine-tuned RT-HF CIR as input features and the actual CIR as the output label. A properly trained DNN model assists in predicting CIR that is much closer to the actual CIR. Hence, the proposed scheme shows a notably reduced gap in BER performance compared to the genie-aided scheme. Moreover, Fig. \ref{fig6} depicts that the proposed DNN model prediction significantly outperforms the baseline MLP DNN model.

To articulate the efficiency of the proposed AI-driven channel twin model, we compute the normalized mean squared error (NMSE) for two cases. Case 1 calculates the NMSE between the actual CIR and the predicted CIR from the fine-tuned RT-HF model, whereas Case 2 computes the NMSE between the actual CIR and the predicted CIR from the proposed data-driven approach. 
\setlength{\textfloatsep}{0pt}
\begin{table}[h]
\centering
\captionsetup{justification=centering}
\caption{NMSE comparison}
\begin{tabular}{|p{2.1cm}|p{2.1cm}|}
\hline
\bfseries{Category} & \bfseries{NMSE [dB]}  \\
\hline
Case 1 & 8.2392  \\
\hline
Case 2 & \textbf{-21.2773}  \\
\hline
\end{tabular}
\label{tab3}
\end{table}
Table \ref{tab3} demonstrates that the NMSE for Case 2 is significantly lower than that for Case 1, highlighting the efficacy of the proposed AI-driven comprehensive channel twin model in accurately computing the Channel Impulse Response (CIR).

\section{Conclusions}
This work proposes a novel approach to design a true physical environment-aware channel twin model that can replicate the channel very close to the real channel. Specifically, we proposed a multi-step fine-tuning approach to rectify CIR generated by RT emulation in order to create the twin of the real channel. First, we propose a two-step tuning approach in RT emulation to generate HF CIR. Then, final tuning with the aid of the proposed AI-driven algorithm can render CIR close to the actual CIR. Simulation results reveal that the proposed precoding scheme can exhibit BER performance close to that of the optimal genie-aided scheme.


\balance
\renewcommand{\baselinestretch}{0.90}

\bibliographystyle{IEEEtran}
\bibliography{Main_File}

\balance

\end{document}